\documentclass[runningheads]{llncs}

\usepackage{amssymb}
\usepackage{pifont}
\newcommand{\cmark}{\ding{51}}%
\newcommand{\xmark}{\ding{55}}%

\usepackage{booktabs}
\usepackage{enumitem}
\usepackage{lipsum}
\usepackage{comment}
\usepackage{helvet}
\usepackage{tabularx}
\usepackage{xurl}
\usepackage{hyperref}

\usepackage{listings}
\usepackage{xcolor}

\definecolor{jsonkey}{HTML}{1A237E}   
\definecolor{jsonstring}{HTML}{388E3C} 
\definecolor{jsonnumber}{HTML}{D32F2F} 

\lstdefinelanguage{json}{
    basicstyle=\ttfamily\scriptsize, 
    breaklines=true,
    showstringspaces=false,
    morestring=[b]",
    stringstyle=\color{jsonstring},
    literate=
     *{0}{{{\color{jsonnumber}0}}}{1}
      {1}{{{\color{jsonnumber}1}}}{1}
      {2}{{{\color{jsonnumber}2}}}{1}
      {3}{{{\color{jsonnumber}3}}}{1}
      {4}{{{\color{jsonnumber}4}}}{1}
      {5}{{{\color{jsonnumber}5}}}{1}
      {6}{{{\color{jsonnumber}6}}}{1}
      {7}{{{\color{jsonnumber}7}}}{1}
      {8}{{{\color{jsonnumber}8}}}{1}
      {9}{{{\color{jsonnumber}9}}}{1}
}

\makeatletter
\lst@AddToHook{Output}{%
    \ifnum\lst@mode=\lst@Pmode
        \def\lst@thestyle{\color{jsonkey}}%
    \fi
}
\makeatother

\usepackage[T1]{fontenc}
\usepackage{graphicx}
\usepackage{xcolor}
\usepackage{multirow}
\usepackage{rotating}
\usepackage{array}

\usepackage{bbding}

\begin{document}
\title{LISP -- A Rich Interaction Dataset and \\Loggable Interactive Search Platform}
%
%
\author{Jana Isabelle Friese\inst{1}\Envelope\orcidID{0009-0005-2483-0476} \and  
Andreas Konstantin Kruff\inst{2}\orcidID{0009-0002-8350-154X} \and
Philipp Schaer\inst{2}\orcidID{0000-0002-8817-4632} \and 
Norbert Fuhr\inst{1}\orcidID{0000-0002-0441-6949} \and
Nicola Ferro\inst{3}\orcidID{0000-0001-9219-6239}
}

\institute{
University of Duisburg-Essen, Germany \\
\email{\{jana.friese,norbert.fuhr\}@uni-due.de}
\and
TH Köln - University of Applied Sciences, Germany \\
\email{\{andreas.kruff,philipp.schaer\}@th-koeln.de} 
\and 
University of Padua, Italy \\
\email{nicola.ferro@unipd.it}
}
\authorrunning{Friese et al.}

\maketitle              
\begin{abstract} We present a reusable dataset and accompanying infrastructure for studying human search behavior in Interactive Information Retrieval (IIR). The dataset combines detailed interaction logs from 61 participants (122 sessions) with user characteristics, including perceptual speed, topic-specific interest, search expertise, and demographic information. To facilitate reproducibility and reuse, we provide a fully documented study setup, a web-based perceptual speed test, and a framework for conducting similar user studies. Our work allows researchers to investigate individual and contextual factors affecting search behavior, and to develop or validate user simulators that account for such variability. We illustrate the dataset’s potential through an illustrative analysis and release all resources as open-access, supporting reproducible research and resource sharing in the IIR community.
\end{abstract}

\section{Motivation}

Resource sharing and re-use are essential for advancing research, ensuring transparency, and fostering community collaboration in Interactive Information Retrieval (IIR) \cite{Gaede2021}.
While a number of IIR studies build on existing datasets, only a small fraction make their own data publicly available, and even fewer share complete experimental setups \cite{Bogers2023}. This mismatch between interest in reusing resources and their availability limits cumulative progress and highlights the need for more accessible, reusable datasets.

A larger pool of reusable resources would directly support IIR research. While user studies provide valuable insights into real-world scenarios beyond system performance \cite{Kelly2009}, human search behavior varies widely, and effect sizes are often small. Still, most IIR studies rely on limited samples, undermining power and reproducibility. 
User simulation offers a scalable alternative \cite{Balog2024}, but click models mostly rely on simplified assumptions about user behavior \cite{Chuklin2015} and, even in advanced frameworks, lack systematic validation against real behavior \cite{Zerhoudi2024}.
Reliable baselines of authentic user behavior can enable reliable comparison and validation of such models, enhancing progress in IIR research.

However, in existing session interaction datasets some crucial information is often missing. Liu and Shah \cite{Liu2019} note that participant variables, which may significantly affect results, are under-reported in IIR studies. Gäde et al. \cite{Gaede2021} identify three main resource types that should be documented and shared to enable effective reuse: (1) \textit{research design}, (2) \textit{infrastructure}, and (3) \textit{data}. In practice, however, sharing typically focuses on the data alone, with limited attention to the other components. In particular, infrastructures—--such as user interfaces and logging frameworks—--are rarely reused but instead redeveloped from scratch, as they are difficult to adapt to new research scenarios \cite{Hall2019}.

Our work addresses this gap by introducing a comprehensive, reusable resource that integrates all three main resource types outlined by Gäde et al. The dataset includes detailed interaction logs enriched with additional participant information (i.e., the \textit{data}), an extensively documented study setup (\textit{design}), and the complete infrastructure for running and adapting the experiment (\textit{infrastructure}). To ensure high reusability, we aligned our materials with the highest of Gäde et al.’s five levels of reusability standards, which require structured and openly documented archival of all three resource types. Because infrastructure often poses the greatest challenge for reuse, we also provide detailed instructions for adapting our setup to different scenarios.\footnote{Resources accessible at: https://github.com/irgroup/LISP\_Dataset\_and\_Platform}

To build the resource, we conducted a user study focusing on perceptual speed and interest in an argument retrieval task, as prior work has shown that both cognitive abilities \cite{AlMaskari2011Effect,Allen1992Cognitive,Brennan2014,Azzopardi2023} and contextual factors \cite{Qu2010,OBrien2020,Vuong2019,Choi2010} shape user behavior. 
We therefore selected perceptual speed and interest as representative dimensions of individual and situational factors. Including these factors makes the dataset suitable not only for studying human search behavior but also for developing and validating user simulators that account for such variability. Alongside the interaction logs, we provide each participant’s perceptual speed scores, topic-specific interest ratings, search expertise, and demographic information.

Our study followed an exploratory design, enabling a wide range of research questions to be addressed. This paper includes an illustrative analysis of the effects of perceptual speed and interest on user behavior, demonstrating how the dataset can be used. The analysis both characterizes the collected data and highlights its potential for developing and evaluating simulators that adapt to different user types or contexts. In doing so, we also discuss the broader implications of these factors for simulation and user-centric evaluation in IIR.

Overall, the dataset and accompanying infrastructure offer the IIR community: (1) a robust baseline of human search behavior accounting for demographics, cognitive abilities, and situational factors; (2) a fully documented, adaptable study environment; and (3) a practical example of designing and sharing reusable resources that adhere to current best-practice standards.

\section{Related Work}

To position our dataset within the current landscape, we conducted a literature review on existing session log resources and their characteristics. As noted by Reimer et al. \cite{Reimer2023}, several large-scale query logs have been collected over the years; however, most focus on individual queries rather than full sessions and do not include click data. Many are also no longer accessible, underscoring the scarcity of suitable datasets and the challenge of long-term reproducibility.

Well-established log datasets, such as the TREC Session Track collections \cite{TREC2014}, remain publicly available and provide high-quality session-based interaction data from controlled lab studies. While being valuable for research, they are limited in size. In contrast, logs collected from real-world search applications enable the creation of much larger datasets. One example is TripClick \cite{Rekabsaz2021}, which offers extensive click and ranking data but lacks any information about users or their underlying information needs. User- and context-related factors, however, strongly influence search behavior. Yet, even in lab-based studies where such information is typically collected, these details are rarely published. 

Detailed information about user characteristics and context is crucial not only for understanding search behavior but also for ensuring reproducibility. In recent years, there has been growing interest in how individual differences and motivational factors shape search behavior. Among the many traits studied, perceptual speed and task-specific interest have frequently been included, highlighting their relevance for modeling user interactions \cite{AlMaskari2011Effect,Brennan2014,Turpin2016,Arguello2019,Azzopardi2023,Kelly2015,Edwards2016,Liu2021,Huang2025}. Without reporting these factors, valuable insights are lost, and the usefulness of datasets is limited. Additional constraints, such as language and domain coverage, further restrict the applicability of existing resources.

Table \ref{tab:related} provides an overview of publicly available session-based interaction log datasets, highlighting characteristics such as number of logged sessions, domain, language, and collection environment. The language column (Lang.) indicates the primary language of the dataset, as many logs—--especially from real-world settings--—contain queries in multiple languages.

\begin{table}[h]
\centering
\caption{Summary of publicly available session-based interaction log datasets. The last row shows our own dataset.}
\label{tab:related}
\begin{tabularx}{\textwidth}{l c X c c c c}
\toprule
Ref & User Profiles & Domain & Lab Setting & Year & \# Logs & Lang.  \\
\midrule
Yandex \cite{Serdyukov2014} & \xmark & Mixed & \xmark & 2011 & 797,867 & en  \\
TREC Session\cite{TREC2014} & \xmark & Mixed & \cmark & 2011--2014 & 1,564 & en \\
TripClick \cite{Rekabsaz2021} & \xmark & Medical & \xmark & 2013--2020 & 1.6M & en  \\
AOL \cite{DBLP:conf/infoscale/PassCT06} & \xmark & Mixed & \xmark & 2006 & 283,207 & en\\
Baidu-ULTR \cite{zou2022} & \xmark & Mixed & \xmark & 2022 & 1.2B  & ch  \\
SoguoQ \cite{Song2021} & \xmark & Mixed & \xmark & 2008 & ~~~~14.1M~~~~ & ch\\
TianGong-ST \cite{DBLP:conf/cikm/ChenMLZM19} & \xmark & Mixed & \xmark & 2015 & 147,155 & ch \\
SUSS \cite{DBLP:conf/ercimdl/MayrK17} & \xmark & Academic & \xmark & 2014--2015 & 484,449 & en/ger\\
\midrule
Own & \cmark & Argument & \cmark & 2025 & 122 & en \\
\bottomrule
\end{tabularx}
\end{table}

While some datasets contain extensive click data, they reveal very little about the users: outside controlled lab settings, information about users’ underlying information needs is missing, and no existing dataset provides comprehensive details on user or contextual factors. 

This aligns with the observations reported by Crasswell et al. \cite{Craswell2020}: privacy concerns and the sensitivity of user-entered data often prevent researchers from sharing their logs, leaving few publicly available click datasets suitable for research. This further underscores the limited availability of resources that support meaningful comparison or reproduction of studies.

The scarcity of datasets is mirrored by the limited availability of shared frameworks for collecting and analyzing interaction data. Several logging frameworks have been proposed \cite{logui,bigbrother,privacyawarelogging,YASBIl}, and initial solutions for RAG systems have emerged \cite{Generative_IR_User_Study_Platform}. However, fully integrated end-to-end frameworks remain rare, and most researchers rely on custom implementations. A notable exception is Podify \cite{podify}, which provides standardized logging and reproducible experiments, but its focus on podcast streaming limits applicability to other document types or search tasks.
This lack of accessible infrastructure aligns with bibliometric findings from Bogers et al. \cite{Bogers2023}, who reported that only a small fraction of CHIIR authors shared research data or infrastructure components, reflecting a broader reproducibility gap in the field.

\section{Resource Design and Development}
\label{sec:designanddevelopment}

To create the dataset, we conducted a user study and collected detailed interaction logs together with extensive participant profiles.
We used a within-subject study design to investigate the effects of topic interest---i.e., a cognitive-emotional relationship between an individual and a topic \cite{Sinnamon2021}---on search behavior. Participants were asked to prepare for writing an opinion essay through an exploratory search, gathering arguments for both sides of a given debate. Each participant worked on two topics: one of high and one of no personal interest, making topic interest the within-subject variable. In addition, perceptual speed (PS)---defined as “the speed in finding figures, making comparisons, and carrying out other very simple tasks involving visual perception” \cite{French1963Manual}---was included as a between-subject variable to assess its impact on search behavior.

The study consisted of three parts: (1) Participants completed a pre-study questionnaire, including the Perceptual Speed Test, both administered remotely. (2) They then attended the main study session, (3) followed by a post-study questionnaire; these latter steps were conducted in person.

\subsection{Dataset and System}
For the experimental setup, we used the Conversational Argument Retrieval dataset from Touché 2020 \cite{Bondarenko2020OverviewOT}, which is based on the args.me corpus of Debate.org threads. The corpus contains 387,606 arguments across 50 TREC-style topics, collected in mid-2019. We selected this corpus because its controversial topics are well suited to elicit user interest, increasing the likelihood of finding topics with different levels of interest. In addition, the relatively short argumentative texts reduce reading time and encourage more frequent system interactions within the ten-minute session limit. Topics that were unlikely to be relatable to the participants (e.g., heavily US-centric issues) were excluded, as well as topics where one side of the debate was underrepresented, resulting in a set of 26 topics. From the corresponding arguments, we removed overly short or uninformative entries (<150 characters) as well as excessively long texts (>3,000 characters) that could distort reading time, leaving 226,468 documents in the collection.

To provide a familiar search experience, we implemented a simple search system using Terrier’s BM25 ranking, allowing users to submit queries, view retrieved arguments, save them as supporting or opposing a stance, and review previously saved documents. Results were shown in pages of ten arguments, each presented with a title and a snippet, expandable to the full text on click. Snippets consisted of the first 200 characters of an argument. As the dataset does not include titles, we generated them using Llama3 (temperature = 0.3) based on the argument text and included them in the GitHub repository with the system. The current topic was displayed alongside the results at all times.

\subsection{Data Collection}

\subsubsection{Participant Recruitment}
For the study, we recruited information science students from a German university. Participation was fully voluntary, with informed consent obtained after participants were briefed on the study purpose, data handling, and their rights to reject the publication of their data. Students were compensated for their participation with course credits. 
All study steps were conducted under full anonymity, using pseudonyms to ensure that no personal data (as defined by the GDPR) was collected. Pseudonyms were further standardized during post-processing to strengthen data protection.

\subsubsection{Questionnaires}

Before the search sessions, participants completed a pre-study questionnaire on demographics, socioeconomic background, online activity, and search engine experience (see \ref{subsec:userprofiledata} for further details). They also rated their interest in the 26 available topics.

At the end of the study, participants had to complete a post-study questionnaire. They were asked whether their stance on the topic had changed during the process, and if so, how it had changed. In addition, participants were asked to rate the difficulty of completing the tasks.
\subsubsection{Perceptual Speed Test}

As part of the pre-study questionnaire, participants completed a perceptual speed test in their own environment to avoid potential lab effects. Several perceptual speed tests have been reviewed in the context of IR \cite{Foulds2020}. Following Azzopardi et al. \cite{Azzopardi2023}, we used a modified Finding A’s test, in which participants identified $\varepsilon$ and ¥ characters, as this method has been shown to be a more accurate test of users actual perceptual speed \cite{Ackerman2000}.

After instructions, participants completed a 30-second practice phase. In the test, participants classified strings of 10\textendash15 non-alphanumeric characters---presented one at a time---as containing both the symbols $\varepsilon$ and ¥ (pressing \textit{j}) or not (pressing \textit{n}). The time limit was two minutes. All participants saw the same fixed order of strings, with a maximum of 500 provided to ensure comparability and prevent ceiling effects.

\subsubsection{Study Procedure}

Based on the pre-study questionnaire, each participant was assigned two topics: one of high and one of no interest to them. To avoid ordering effects, half of the participants started with the high-interest and half with the no-interest topic. Topics were assigned using a heuristic pseudo-random approach to ensure a balanced coverage across participants, preventing any single topic from dominating either interest category. The task was identical for both topics. After an introduction explaining the procedure and task, the participants started the study on the provided computers. Each task had a ten-minute time limit. Upon completing the first task, they were automatically directed to the second. Before each task, participants were shown the full topic and task description. After completion, they could optionally explain why they ended their search before the time limit. These explanations are included in the dataset in their original form to avoid bias; for responses written in German, a machine-translated version is also provided (based on Google Translate API). Upon finishing both tasks, participants proceeded to the post-study questionnaire.

\section{Resource Description}
Building on the study design described in Section \ref{sec:designanddevelopment}, this section documents the resources we release: (1) the interaction log dataset, (2) the loggable interactive search platform, and (3) the user profile data (demographics, perceptual speed, and search expertise). A more comprehensive documentation is available on the project website\footnote{https://irgroup.github.io/LISP\_Dataset\_and\_Platform/} to support reproducibility and adhere to the Level-5 reusability standard.  

\subsection{Interaction Log Dataset}
\label{subsec:logs}

The interaction dataset collected in the user study comprises 122 session log files, evenly divided between high-and no-interest topics. Each interaction contains an interaction type, a timestamp, and session identifier.

Events captured include query submissions and reformulations, document interactions (such as expanding or collapsing views, and selecting stances), navigating between result pages, and task completion actions. Each document entry is associated with its rank, retrieval score, and length. 
Stopping decisions were explicitly logged, and participants could provide a reason for ending their search early. Although each task was designed for a ten-minute duration, this limit was not hard-coded, so some sessions slightly exceeded this timeframe.

\subsection{lisp - A Loggable Interactive Search Platform}

For the study, we developed a loggable interactive search platform called lisp. While originally designed for argument retrieval tasks, the platform can be easily adapted to other search scenarios.

\begin{figure}[b]
\centering
\includegraphics[width=0.82\textwidth]{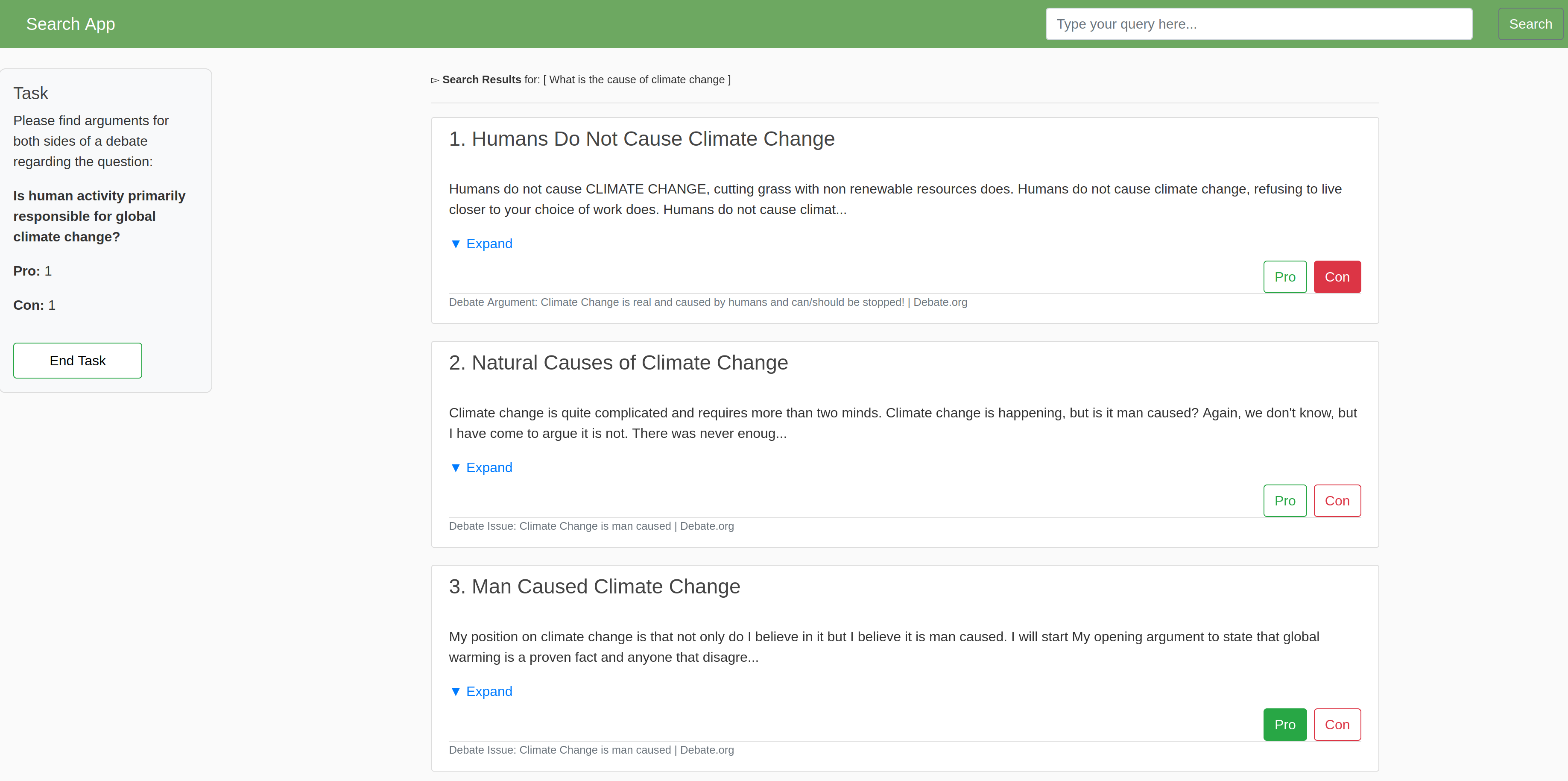}
\caption{Screenshot of the search interface of lisp} \label{img:interface}
\end{figure}

The interface provides a typical search experience: participants can enter queries, browse retrieved arguments, navigate between pages, expand snippets to view the full text, and save documents as supporting or opposing a stance. A side panel summarizes the task and topic, and tracks the number of labeled documents (Figure \ref{img:interface}). At the end of each session, a pop-up window presents an overview of all saved documents with their assigned labels and titles, with stance indicated through color-coding (i.e., green for supporting, red for opposing).

For each session, all user interactions described in \ref{subsec:logs} are logged, and a logfile is created. Logging begins when a task is started and ends when the participant confirms completion. Logfiles are named using the participant ID, task number, and timestamp.

In the current workflow, participants first see a welcome page, then enter their username, and proceed through two successive search tasks before finishing on a page linking to the post-study questionnaire. 

The platform is designed for easy adaptation: elements such as buttons, result presentation, logging output, and datasets can all be customized with minimal changes. Detailed instructions are provided in the repository.

The repository additionally includes the implementation of the Perceptual Speed Test. Running the test requires a MySQL database for storing results, and the GitHub README provides instructions on the necessary implementation changes. A demo version\footnote{https://andykruff.github.io/demo-ps-test/} of the Perceptual Speed Test is also provided, which allows for testing the application without setting up a database.

\subsection{User Profile Data} 
\label{subsec:userprofiledata}

\begin{figure}[h!]
    \centering
    \includegraphics[width=.91\textwidth]{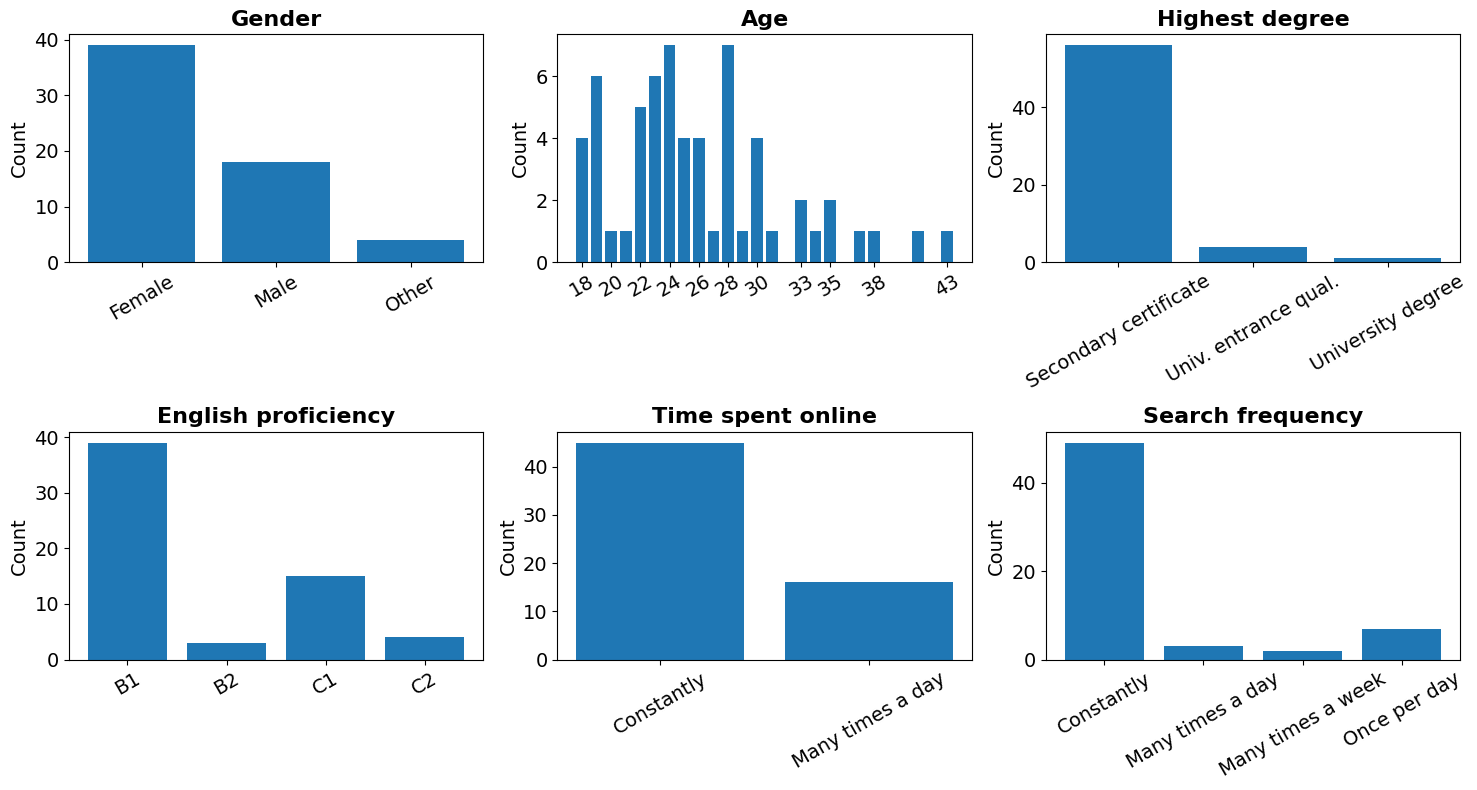}
\caption{Demographic and experience profile of the user sample (N = 61). Answer options that were not selected are not displayed. \\ \label{img:prestudy}}
    \includegraphics[width=0.7\textwidth]{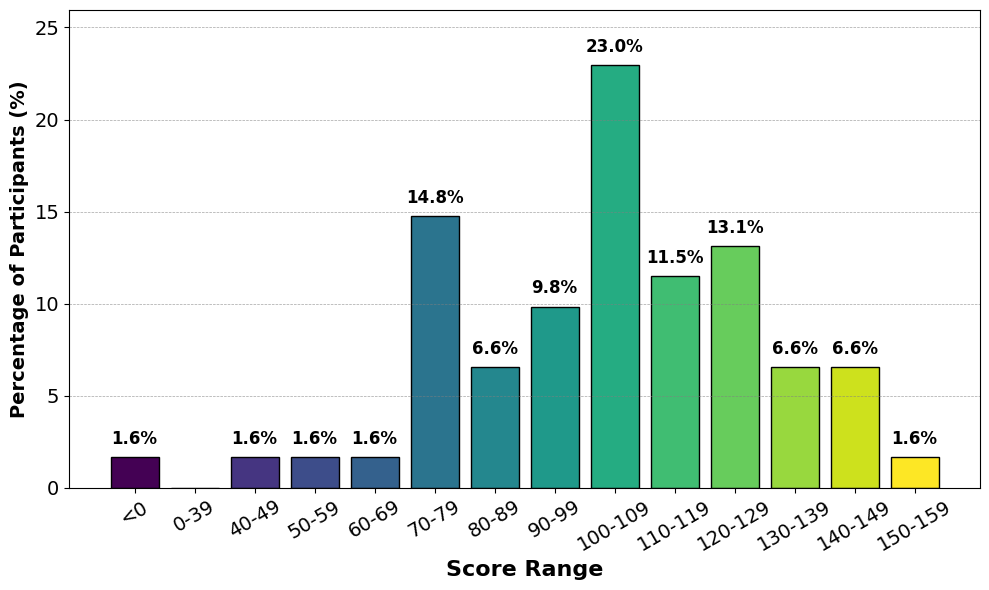}
\caption{Distribution of perceptual speed scores among participants (N = 61)\label{img:pstest} }

\end{figure}

In addition to interaction logs, the dataset includes user profiles with demographics, search experience, and perceptual speed scores. Figure~\ref{img:prestudy} summarizes the demographic and experience data. As all participants were recruited from the same university course, the sample is relatively homogeneous.

Perceptual speed results (Figure~\ref{img:pstest}) show an average score of 101.66 with a median of 103 and a standard deviation of 28.5, indicating substantial variability within the group. The dataset also reports detailed response patterns: the number of \textit{j}- and \textit{n}-key presses (indicating both symbols were perceived or not) and the correctness of these responses. Failure rates were low, with 1.72\% false positives and 1.32\% false negatives, and an overall failure rate of 1.46\%.

\section{Influence of Perceptual Speed and Interest on Search Behaviors}

To illustrate the practical value of the dataset collected through our framework, we present an initial case study on the questions how perceptual speed and the level of interest influences the observed search behaviour.

While the main contribution of this paper is the dataset itself, we provide this case study as an illustrative analysis to demonstrate the potential use cases of our research resource.  
We address the following research questions:
\begin{enumerate}[label=\bfseries RQ\arabic*:,leftmargin=*,labelindent=1em]
    \item How does perceptual speed influence the observed search behaviors of users?
    \item How does their level of interest in a topic influence the observed search behaviors of users?
\end{enumerate}

\subsection{Outcome Measures}

To comprehensively assess search behavior, we considered interactions and times. Most interactions were extracted directly from the logfiles, while interaction durations were estimated from timestamps (e.g., document viewing time was measured from click to marking or the next document click). Interactions considered were queries issued, documents clicked or marked, pages viewed, and snippets viewed, both per session and per query; times included session duration, time per query formulation, per query, per document, and per snippet.

\begin{figure}
\centering
\includegraphics[width=0.75\textwidth]{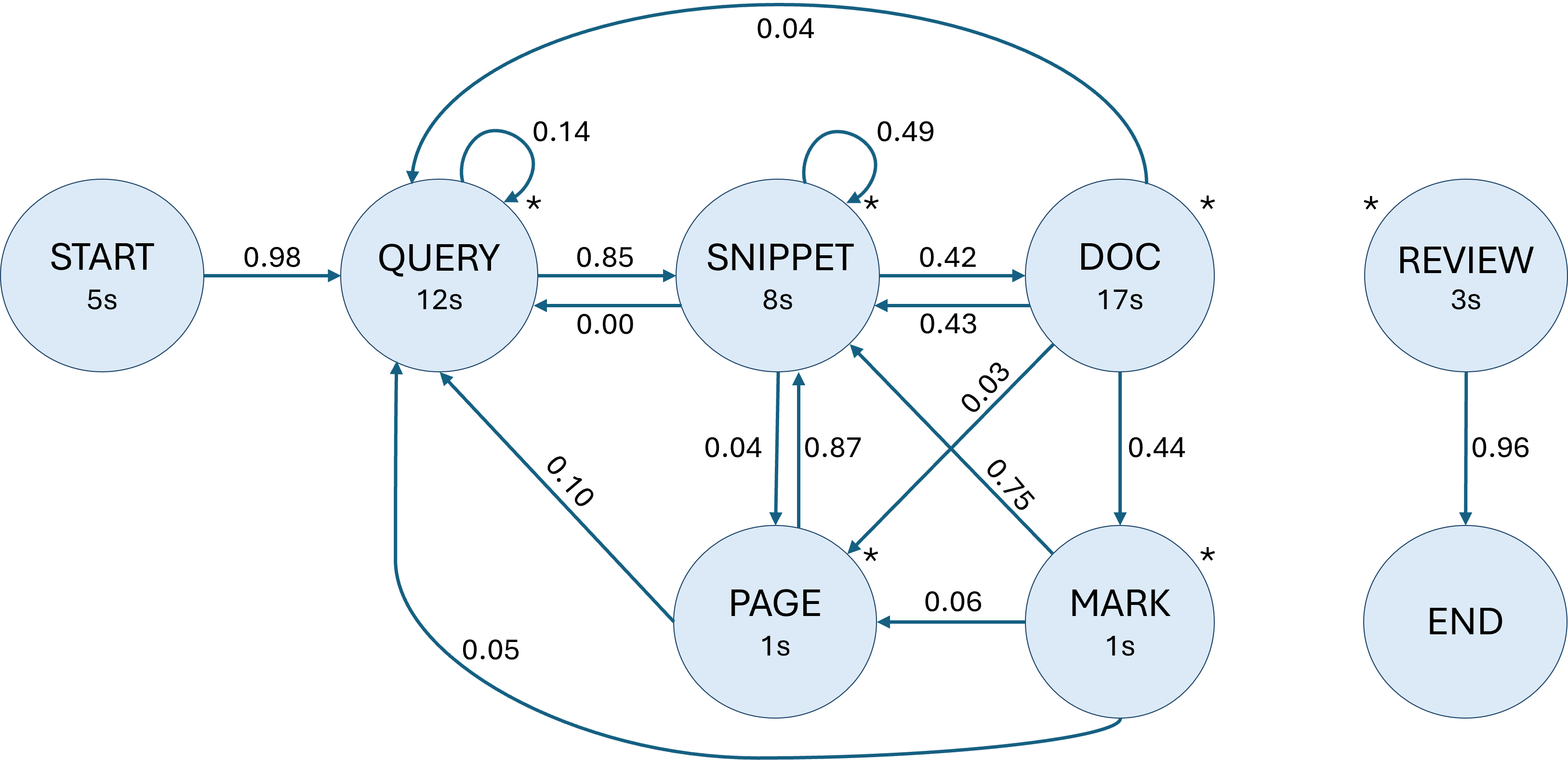}
\caption{Markov model of user interactions. Transitions from states marked with * to {\sffamily REVIEW} and from {\sffamily REVIEW}  back to those states were omitted for clarity. Transition probabilities indicated at the arrows and mean interaction times per state are averaged across all sessions. As {\sffamily MARK} and {\sffamily PAGE} have no actual duration, they are assigned a nominal value of 1 second for completeness.} \label{img:model}
\end{figure}

Since common measures do not capture the full search process, we also built Markov models from the interaction sequences to compare entire search processes across groups. Figure \ref{img:model} shows the resulting Markov model based on all sessions.

\subsection{Analysis}
We used the Wilcoxon signed-rank test as a non-parametric alternative to the paired t-test to examine differences between conditions of the within-subject variable (topic interest), since the data were not normally distributed. Differences in the between-subject variable (PS) were analyzed using the Mann–Whitney U test, the non-parametric equivalent of the independent-samples t-test. Participants were divided into high- and low-PS groups based on the sample median, following previous work \cite{Arguello2019,Turpin2016,Azzopardi2023,AlMaskari2011Effect,Brennan2014}; the median value was included in the higher group to ensure more balanced group sizes. To control for multiple testing, we applied the Benjamini–Hochberg correction.

Regarding the Markov models, we used the Frobenius norm to assess overall model similarity, while using the Jensen–Shannon divergence and the Kolmogorov–Smirnov test to assess differences at the state level.

\subsection{Results}
\subsubsection{Perceptual Speed}
Comparing users with high versus low perceptual speed revealed only minor differences (see Table \ref{tab:results}). On average, high-PS users spent 2.94 seconds longer on each document and examined 9.89 fewer snippets per session. However, none of these differences reached statistical significance.

The Markov models also showed minimal variation between groups (Frobenius norm = 0.18), while state-level comparisons yielded very low Jensen–Shannon divergences (0.00\textendash0.13). Likewise, Kolmogorov–Smirnov tests indicated no significant differences (all $p > 0.98$) between transition probability distributions. 

\begin{table}[t]
\centering
\caption{Mean values and statistical results for all behavioral measures by Interest (\textit{high int.} vs. \textit{no int.}) and Perceptual Speed (\textit{high PS} vs. \textit{low PS}). For Interest, the mean within-participant differences (MD) between high- and no-interest topics are additionally reported. Statistically significant results ($p < 0.05$) are indicated by an asterisk (*).}

\begin{tabular}{|c| r | >{\centering}m{1.2cm} | >{\centering}m{1.2cm} | >{\centering}m{1cm} | >{\centering}m{1.3cm} !{\vrule width 1.2pt} >{\centering}m{1.2cm} | >{\centering}m{1.2cm} | >{\centering\arraybackslash}m{1.3cm} |}
\cline{3-9}
\multicolumn{2}{r|}{}& \multicolumn{4}{c!{\vrule width 1.2pt}}{Interest} & \multicolumn{3}{c|}{Perceptual Speed} \\
\hline
\multicolumn{2}{|r|}{Measure} & high int.  & no int.  & MD & $p$ (corr.) & high PS & low PS & $p$ (corr.) \\
\noalign{\hrule height 1.2pt}
\multirow{5}{*}{\rotatebox{90}{$\#.../session$}}&Queries & 3.82  & 2.95  & 0.87 & 0.037* & 3.35  & 3.42  & 0.990 \\
&Documents & 23.02  & 21.52  & 1.49 & 0.458 & 22.29  & 22.25  & 0.823 \\
&Marked docs & 12.36  & 14.13  & -1.77 & 0.110 & 12.82  & 13.68  & 0.208 \\
&Snippets & 51.93 & 47.85 & 4.08 & 0.458 & 45.03  & 54.92  & 0.140 \\
&Pages & 7.16  & 6.10  & 1.07 & 0.110 & 6.00  & 7.28  & 0.188 \\
\hline
\multirow{4}{*}{\rotatebox{90}{$\#.../query$}} &Documents & 10.01  & 12.24  & -2.23 & 0.091 & 11.16  & 11.08  & 0.847 \\
&Marked docs & 5.16  & 7.61  & -2.44 & 0.028* & 5.78  & 7.01  & 0.723 \\
&Snippets & 18.44  & 25.32  & -6.88 & 0.042* & 19.37  & 24.48  & 0.268 \\
&Pages & 2.28  & 2.78  & -0.50 & 0.090 & 2.23  & 2.83  & 0.182 \\
\hline
\multirow{5}{*}{\rotatebox{90}{$time~per...$}}&Session & 662.88  & 653.12  & 9.76 & 0.458 & 665.98  & 649.75  & 0.188 \\
&Query & 276.78  & 360.32  & -83.54 & 0.037* & 320.39  & 316.64  & 0.823 \\
&Q.Formulation & 11.62  & 12.36  & -0.74 & 0.580 & 11.93  & 12.04  & 0.823 \\
&Snippet & 8.17  & 7.90  & 0.27 & 0.580 & 7.63  & 8.45  & 0.939 \\
&Document & 16.62  & 18.23  & -1.61 & 0.860  & 18.87  & 15.93  & 0.140 \\
\hline
\end{tabular}
\label{tab:results}
\end{table}

\subsubsection{Interest}

Aggregated over entire sessions, the only significant difference for interest was in the number of queries: participants submitted more queries for topics of personal interest. At a finer level, however, clear differences emerged: users spent less time per query, clicked and marked fewer documents, and viewed fewer pages and snippets, with time per query and the numbers of viewed snippets and marked documents reaching significance (see Table \ref{tab:results}).

The Markov models revealed no notable differences across conditions: overall, a Frobenius norm of 0.10 indicates highly similar behavior, while state-level comparisons showed minimal differences (Jensen–Shannon divergence = 0.02\textendash0.11) and Kolmogorov–Smirnov tests confirmed no significant effects (all $p > 0.98$).
\paragraph{Post-Study questionnaire}
\label{par:post_study}

We analyzed post-questionnaire responses to examine how topic interest affected stance changes, certainty, and perceived task difficulty. Users rarely changed stance (5 for high-interest vs. 7 for no-interest topics), and interest did not affect task difficulty (2.66 vs. 2.87). Nearly one-third of participants reported changes in certainty: for no-interest topics, increases and decreases were balanced (9 vs. 8), whereas for high-interest topics, more participants became more certain (13) than less certain (3).

\subsection{Discussion}
\subsubsection{Perceptual Speed} 
None of the observed trends for perceptual speed reached significance, likely due to the exploratory nature of our study and stricter corrections for multiple comparisons. Nevertheless, the trends—longer document viewing times and fewer viewed snippets for high-PS users—align with Azzopardi et al. \cite{Azzopardi2023}, despite contradicting the expectation that higher perceptual speed leads to shorter viewing times. Overall, the results suggest that perceptual speed does influence search behavior, even if not statistically conclusive here.

\subsubsection{Interest}
Users formulated more queries for topics of interest but spent less time per query, clicked fewer documents, and marked fewer arguments. Assuming greater prior knowledge, they may have approached searches with higher expectations, which can influence behavior \cite{Wang2023}. However, this contrasts with Edwards et al. \cite{Edwards2016}, who found that task interest affected engagement but not observable behavior. A likely explanation for this discrepancy is the task design:
Edwards et al. examined information-seeking tasks requiring the identification and evaluation of multiple options from neutral sources, whereas our study focused on gathering subjective arguments for controversial topics. 
As topics rated as interesting likely coincide with strong pre-existing opinions, cognitive biases like the confirmation bias were likely more present \cite{Nickerson1998,Azzopardi2021}. Though they were instructed to consider both sides, interested users may have actively sought evidence confirming their beliefs, leading to more focused and less exploratory behavior. In contrast, the tasks in Edwards et al. required an inherently exploratory approach.
These results indicate that the impact of interest on search behavior depends strongly on task type and cognitive framing. Distinguishing between cognitive and emotional involvement may therefore be crucial for understanding interest effects in interactive IR. 

\subsection{Limitations}
Although the dataset provides rich and detailed information, it is limited in size, which means that some results are not statistically conclusive, particularly for subtle effects with small effect sizes. Consequently, the findings may not generalize beyond the specific setting examined. Moreover, the focus on a single domain further restricts the scope of applicability of the results.

\subsection{Conclusion}
Overall, the findings suggest that user characteristics, as well as the context of a search influence search behavior and should be accounted for in user models and simulators.
At the same time, the Markov model results illustrate the limits of aggregated process representations: Although interest effects were more pronounced than those of perceptual speed, the Markov models for high- and no-interest sessions were more similar, since the models average over entire sessions while the interest effects primarily occurred at the query-level. Because querying is relatively infrequent compared to other interactions, the observed differences in query behavior had little impact on the overall transition probabilities. In addition, opposing tendencies within a session may balance out. This highlights the problem of oversimplifying user behavior for simulations: without considering search context---such as the current query, document rank, or evolving task state---important dynamics are lost. For user models and simulators to be realistic, they need to capture not only aggregate session patterns but also contextual dependencies that shape search as it unfolds.

\section{Outlook}

Beyond potential insights into how interest and perceptual speed shape search behavior, the presented resource opens up multiple avenues for future research. The dataset can support both empirical and modeling work---for instance, in validating and refining user simulators that account for cognitive traits and situational factors, or in developing models that incorporate retrieval scores and document characteristics to explain user interactions. The detailed log files also allow for analyses such as inter-rater agreement on marked arguments across participants and topics, offering new perspectives on subjectivity and consistency in argument relevance.

From a practical perspective, the data can also inform the design of user-adaptive search systems. By revealing how users with different cognitive profiles or levels of interest interact with search results, the dataset can be used to study personalization strategies, interface adaptations, or ranking approaches aimed at supporting engagement and efficient information exploration in argumentative or opinion-driven search scenarios.

In addition, the adaptable study setup and open-source infrastructure provide a foundation for conducting new user studies with different contexts. Together, these resources enable cumulative IIR research that bridges human behavior, simulation, and system evaluation.

\begin{credits}
\subsubsection{\ackname}
This work was funded by the Deutsche Forschungsgemeinschaft (DFG, German
Research Foundation) - 509543643. 

\subsubsection{Disclosure of Interests.}
The authors have no competing interests to declare that are relevant to the content of this article.

\end{credits}

\bibliographystyle{splncs04}
\bibliography{user_study.bib}

@inproceedings{Bondarenko2020OverviewOT,
  title={Overview of Touch{\'e} 2020: Argument Retrieval},
  author={Alexander Bondarenko and Maik Fr{\"o}be and Meriem Beloucif and Lukas Gienapp and Yamen Ajjour and Alexander Panchenko and Christian Biemann and Benno Stein and Henning Wachsmuth and Martin Potthast and Matthias Hagen},
  booktitle={Conference and Labs of the Evaluation Forum},
  year={2020},
  url={https://api.semanticscholar.org/CorpusID:225073856}
}

@article{AlMaskari2011Effect,
title = {The effect of user characteristics on search effectiveness in information retrieval},
journal = {Information Processing \& Management},
volume = {47},
number = {5},
pages = {719-729},
year = {2011},
note = {Managing and Mining Multilingual Documents},
issn = {0306-4573},
doi = {https://doi.org/10.1016/j.ipm.2011.03.002},
url = {https://www.sciencedirect.com/science/article/pii/S030645731100029X},
author = {Azzah Al-Maskari and Mark Sanderson},
}

@inproceedings{French1963Manual,
  title={Manual for kit of factor-referenced cognitive tests},
  author={Ruth B. Ekstrom and John W. French and Harry H. Harman},
  year={1976},
  url={https://api.semanticscholar.org/CorpusID:141329865}
}

@inproceedings{Foulds2020,
author = {Foulds, Olivia and Azzopardi, Leif and Halvey, Martin},
title = {Reflecting upon Perceptual Speed Tests in Information Retrieval: Limitations, Challenges, and Recommendations},
year = {2020},
isbn = {9781450368926},
publisher = {Association for Computing Machinery},
address = {New York, NY, USA},
url = {https://doi.org/10.1145/3343413.3377982},
doi = {10.1145/3343413.3377982},
booktitle = {Proceedings of the 2020 Conference on Human Information Interaction and Retrieval},
pages = {234–242},
numpages = {9},
location = {Vancouver BC, Canada},
series = {CHIIR '20}
}

@inproceedings{Allen1992Cognitive,
author = {Allen, Bryce},
title = {Cognitive differences in end user searching of a CD-ROM index},
year = {1992},
isbn = {0897915232},
publisher = {Association for Computing Machinery},
address = {New York, NY, USA},
url = {https://doi.org/10.1145/133160.133212},
doi = {10.1145/133160.133212},
booktitle = {Proceedings of the 15th Annual International ACM SIGIR Conference on Research and Development in Information Retrieval},
pages = {298–309},
numpages = {12},
location = {Copenhagen, Denmark},
series = {SIGIR '92}
}

@inproceedings{Azzopardi2023,
author = {Azzopardi, Leif and Maxwell, David and Halvey, Martin and Hauff, Claudia},
title = {Driven to Distraction: Examining the Influence of Distractors on Search Behaviours, Performance and Experience},
year = {2023},
isbn = {9798400700354},
publisher = {Association for Computing Machinery},
address = {New York, NY, USA},
url = {https://doi.org/10.1145/3576840.3578298},
doi = {10.1145/3576840.3578298},
booktitle = {Proceedings of the 2023 Conference on Human Information Interaction and Retrieval},
pages = {83–94},
numpages = {12},
location = {Austin, TX, USA},
series = {CHIIR '23}
}

@article{Ackerman2000,
  author    = {Phillip L. Ackerman and A. T. Cianciolo},
  title     = {Cognitive, perceptual-speed, and psychomotor determinants of individual differences during skill acquisition},
  journal   = {Journal of Experimental Psychology: Applied},
  year      = {2000},
  volume    = {6},
  number    = {4},
  pages     = {259--290},
  doi       = {10.1037//1076-898x.6.4.259},
  month     = {Dec},
  publisher = {American Psychological Association},
  address   = {United States}
}

@article{Wang2023,
title = {Investigating the role of in-situ user expectations in Web search},
journal = {Information Processing \& Management},
volume = {60},
number = {3},
pages = {103300},
year = {2023},
issn = {0306-4573},
doi = {https://doi.org/10.1016/j.ipm.2023.103300},
url = {https://www.sciencedirect.com/science/article/pii/S0306457323000377},
author = {Ben Wang and Jiqun Liu}
}

@inproceedings{Sinnamon2021,
author = {Sinnamon, Luanne and Tamim, Limor and Dodson, Samuel and O'Brien, Heather L.},
title = {Rethinking Interest in Studies of Interactive Information Retrieval},
year = {2021},
isbn = {9781450380553},
publisher = {Association for Computing Machinery},
address = {New York, NY, USA},
url = {https://doi.org/10.1145/3406522.3446031},
doi = {10.1145/3406522.3446031},
booktitle = {Proceedings of the 2021 Conference on Human Information Interaction and Retrieval},
pages = {39–49},
numpages = {11},
location = {Canberra ACT, Australia},
series = {CHIIR '21}
}

@inproceedings{Qu2010,
author = {Qu, Peng and Liu, Chang and Lai, Maosheng},
title = {The effect of task type and topic familiarity on information search behaviors},
year = {2010},
isbn = {9781450302470},
publisher = {Association for Computing Machinery},
address = {New York, NY, USA},
url = {https://doi.org/10.1145/1840784.1840841},
doi = {10.1145/1840784.1840841},
booktitle = {Proceedings of the Third Symposium on Information Interaction in Context},
pages = {371–376},
numpages = {6},
location = {New Brunswick, New Jersey, USA},
series = {IIiX '10}
}

@article{OBrien2020,
title = {An empirical study of interest, task complexity, and search behaviour on user engagement},
journal = {Information Processing \& Management},
volume = {57},
number = {3},
pages = {102226},
year = {2020},
issn = {0306-4573},
doi = {https://doi.org/10.1016/j.ipm.2020.102226},
url = {https://www.sciencedirect.com/science/article/pii/S0306457319301591},
author = {Heather L. O'Brien and Jaime Arguello and Rob Capra}
}

@article{Choi2010,
author = {Choi, Youngok},
title = {Effects of contextual factors on image searching on the Web},
journal = {Journal of the American Society for Information Science and Technology},
volume = {61},
number = {10},
pages = {2011-2028},
doi = {https://doi.org/10.1002/asi.21386},
url = {https://onlinelibrary.wiley.com/doi/abs/10.1002/asi.21386},
eprint = {https://onlinelibrary.wiley.com/doi/pdf/10.1002/asi.21386},
year = {2010}
}

@article{Vuong2019,
author = {Vuong, Tung and Saastamoinen, Miamaria and Jacucci, Giulio and Ruotsalo, Tuukka},
title = {Understanding user behavior in naturalistic information search tasks},
journal = {Journal of the Association for Information Science and Technology},
volume = {70},
number = {11},
pages = {1248-1261},
doi = {https://doi.org/10.1002/asi.24201},
url = {https://asistdl.onlinelibrary.wiley.com/doi/abs/10.1002/asi.24201},
eprint = {https://asistdl.onlinelibrary.wiley.com/doi/pdf/10.1002/asi.24201},
year = {2019}
}

@inproceedings{Zerhoudi2024,
  author    = {S. Zerhoudi and M. Granitzer},
  title     = {Beyond Conventional Metrics: Assessing User Simulators in Information Retrieval},
  booktitle = {Proceedings of the 14th Italian Information Retrieval Workshop},
  volume    = {3802},
  pages     = {3--12},
  year      = {2024}
}

@inproceedings{TREC2014,
  author    = {B. Carterette and E. Kanoulas and M. M. Hall and P. D. Clough},
  title     = {Overview of the TREC 2014 Session Track},
  booktitle = {Proceedings of The Twenty-Third Text Retrieval Conference (TREC 2014)},
  editor    = {E. M. Voorhees and A. Ellis},
  volume    = {500-308},
  pages     = {1--12},
  year      = {2014},
  publisher = {National Institute of Standards and Technology (NIST)},
  url       = {http://trec.nist.gov/pubs/trec23/papers/overview-session.pdf}
}

@article{Kelly2009,
author = {Kelly, Diane},
title = {Methods for Evaluating Interactive Information Retrieval Systems with Users},
year = {2009},
issue_date = {January 2009},
publisher = {Now Publishers Inc.},
address = {Hanover, MA, USA},
volume = {3},
number = {1—2},
issn = {1554-0669},
url = {https://doi.org/10.1561/1500000012},
doi = {10.1561/1500000012},
journal = {Found. Trends Inf. Retr.},
month = jan,
pages = {1–224},
numpages = {224}
}

@book{Liu2019,
  author    = {J. Liu and C. Shah},
  title     = {Interactive IR User Study: Design, Evaluation, and Reporting},
  publisher = {Morgan \& Claypool Publishers},
  year      = {2019}
}

@article{Balog2024,
url = {http://dx.doi.org/10.1561/1500000098},
year = {2024},
volume = {18},
journal = {Foundations and Trends® in Information Retrieval},
title = {User Simulation for Evaluating Information Access Systems},
doi = {10.1561/1500000098},
issn = {1554-0669},
number = {1-2},
pages = {1-261},
author = {Krisztian Balog and ChengXiang Zhai}
}

@book{Chuklin2015,
  author    = {A. Chuklin and I. Markov and M. de Rijke},
  title     = {Click Models for Web Search},
  series    = {Synthesis Lectures on Information Concepts, Retrieval, and Services},
  publisher = {Morgan \& Claypool Publishers},
  year      = {2015}
}

@inproceedings{Generative_IR_User_Study_Platform,
author = {Liang, Yidong and Wu, Zhijing and He, Yuchen and Liang, Fengming and Liu, Kexin and Mao, Jiaxin},
title = {A Flexible User Study Platform for Generative Information Retrieval},
year = {2025},
isbn = {9798400715921},
publisher = {Association for Computing Machinery},
address = {New York, NY, USA},
url = {https://doi.org/10.1145/3726302.3730140},
doi = {10.1145/3726302.3730140},
booktitle = {Proceedings of the 48th International ACM SIGIR Conference on Research and Development in Information Retrieval},
pages = {4066–4070},
numpages = {5},
location = {Padua, Italy},
series = {SIGIR '25}
}

@inproceedings{YASBIl,
author = {Bhattacharya, Nilavra and Gwizdka, Jacek},
title = {YASBIL: Yet Another Search Behaviour (and) Interaction Logger},
year = {2021},
isbn = {9781450380379},
publisher = {Association for Computing Machinery},
address = {New York, NY, USA},
url = {https://doi.org/10.1145/3404835.3462800},
doi = {10.1145/3404835.3462800},
booktitle = {Proceedings of the 44th International ACM SIGIR Conference on Research and Development in Information Retrieval},
pages = {2585–2589},
numpages = {5},
location = {Virtual Event, Canada},
series = {SIGIR '21}
}

@inproceedings{privacyawarelogging,
author = {Li, Hanyu and Lu, Hongyu and Huang, Songhao and Ma, Weizhi and Zhang, Min and Liu, Yiqun and Ma, Shaoping},
title = {Privacy-Aware Remote Information Retrieval User Experiments Logging Tool},
year = {2021},
isbn = {9781450380379},
publisher = {Association for Computing Machinery},
address = {New York, NY, USA},
url = {https://doi.org/10.1145/3404835.3462793},
doi = {10.1145/3404835.3462793},
booktitle = {Proceedings of the 44th International ACM SIGIR Conference on Research and Development in Information Retrieval},
pages = {2615–2619},
numpages = {5},
location = {Virtual Event, Canada},
series = {SIGIR '21}
}

@inproceedings{bigbrother,
author = {Scells, Harrisen and Jimmy and Zuccon, Guido},
title = {Big Brother: A Drop-In Website Interaction Logging Service},
year = {2021},
isbn = {9781450380379},
publisher = {Association for Computing Machinery},
address = {New York, NY, USA},
url = {https://doi.org/10.1145/3404835.3462781},
doi = {10.1145/3404835.3462781},
booktitle = {Proceedings of the 44th International ACM SIGIR Conference on Research and Development in Information Retrieval},
pages = {2590–2594},
numpages = {5},
location = {Virtual Event, Canada},
series = {SIGIR '21}
}

@InProceedings{logui,
author="Maxwell, David
and Hauff, Claudia",
editor="Hiemstra, Djoerd
and Moens, Marie-Francine
and Mothe, Josiane
and Perego, Raffaele
and Potthast, Martin
and Sebastiani, Fabrizio",
title="LogUI: Contemporary Logging Infrastructure for Web-Based Experiments",
booktitle="Advances in  Information Retrieval",
year="2021",
publisher="Springer International Publishing",
address="Cham",
pages="525--530",
isbn="978-3-030-72240-1"
}

@inproceedings{podify,
author = {Meggetto, Francesco and Moshfeghi, Yashar},
title = {Podify: A Podcast Streaming Platform with Automatic Logging of User Behaviour for Academic Research},
year = {2023},
isbn = {9781450394086},
publisher = {Association for Computing Machinery},
address = {New York, NY, USA},
url = {https://doi.org/10.1145/3539618.3591824},
doi = {10.1145/3539618.3591824},
booktitle = {Proceedings of the 46th International ACM SIGIR Conference on Research and Development in Information Retrieval},
pages = {3215–3219},
numpages = {5},
location = {Taipei, Taiwan},
series = {SIGIR '23}
}

@inproceedings{Gaede2021,
author = {G\"{a}de, Maria and Koolen, Marijn and Hall, Mark and Bogers, Toine and Petras, Vivien},
title = {A Manifesto on Resource Re-Use in Interactive Information Retrieval},
year = {2021},
isbn = {9781450380553},
publisher = {Association for Computing Machinery},
address = {New York, NY, USA},
url = {https://doi.org/10.1145/3406522.3446056},
doi = {10.1145/3406522.3446056},
booktitle = {Proceedings of the 2021 Conference on Human Information Interaction and Retrieval},
pages = {141–149},
numpages = {9},
location = {Canberra ACT, Australia},
series = {CHIIR '21}
}

@inproceedings{Bogers2023,
author = {Bogers, Toine and G\"{a}de, Maria and Hall, Mark Michael and Koolen, Marijn and Petras, Vivien and Larsen, Birger},
title = {How we Work, Share, and Re-use at CHIIR},
year = {2023},
isbn = {9798400700354},
publisher = {Association for Computing Machinery},
address = {New York, NY, USA},
url = {https://doi.org/10.1145/3576840.3578305},
doi = {10.1145/3576840.3578305},
booktitle = {Proceedings of the 2023 Conference on Human Information Interaction and Retrieval},
pages = {351–356},
numpages = {6},
location = {Austin, TX, USA},
series = {CHIIR '23}
}

@inproceedings{Hall2019,
  author    = {M. M. Hall},
  title     = {To re-use is to re-write: Experiences with re-using IIR experiment software},
  booktitle = {CEUR Workshop Proceedings},
  volume    = {2337},
  pages     = {19--23},
  year      = {2019}
}

@inproceedings{Brennan2014,
author = {Brennan, Kathy and Kelly, Diane and Arguello, Jaime},
title = {The effect of cognitive abilities on information search for tasks of varying levels of complexity},
year = {2014},
isbn = {9781450329767},
publisher = {Association for Computing Machinery},
address = {New York, NY, USA},
url = {https://doi.org/10.1145/2637002.2637022},
doi = {10.1145/2637002.2637022},
booktitle = {Proceedings of the 5th Information Interaction in Context Symposium},
pages = {165–174},
numpages = {10},
location = {Regensburg, Germany},
series = {IIiX '14}
}

@inproceedings{Turpin2016,
author = {Turpin, Lauren and Kelly, Diane and Arguello, Jaime},
title = {To Blend or Not to Blend? Perceptual Speed, Visual Memory and Aggregated Search},
year = {2016},
isbn = {9781450340694},
publisher = {Association for Computing Machinery},
address = {New York, NY, USA},
url = {https://doi.org/10.1145/2911451.2914739},
doi = {10.1145/2911451.2914739},
booktitle = {Proceedings of the 39th International ACM SIGIR Conference on Research and Development in Information Retrieval},
pages = {1021–1024},
numpages = {4},
location = {Pisa, Italy},
series = {SIGIR '16}
}

@article{Arguello2019,
author = {Arguello, Jaime and Choi, Bogeum},
title = {The Effects of Working Memory, Perceptual Speed, and Inhibition in Aggregated Search},
year = {2019},
issue_date = {July 2019},
publisher = {Association for Computing Machinery},
address = {New York, NY, USA},
volume = {37},
number = {3},
issn = {1046-8188},
url = {https://doi.org/10.1145/3322128},
doi = {10.1145/3322128},
journal = {ACM Trans. Inf. Syst.},
month = may,
articleno = {36},
numpages = {34}
}

@inproceedings{Kelly2015,
author = {Kelly, Diane and Arguello, Jaime and Edwards, Ashlee and Wu, Wan-ching},
title = {Development and Evaluation of Search Tasks for IIR Experiments using a Cognitive Complexity Framework},
year = {2015},
isbn = {9781450338332},
publisher = {Association for Computing Machinery},
address = {New York, NY, USA},
url = {https://doi.org/10.1145/2808194.2809465},
doi = {10.1145/2808194.2809465},
booktitle = {Proceedings of the 2015 International Conference on The Theory of Information Retrieval},
pages = {101–110},
numpages = {10},
location = {Northampton, Massachusetts, USA},
series = {ICTIR '15}
}

@inproceedings{Edwards2016,
author = {Edwards, Ashlee and Kelly, Diane},
title = {How does Interest in a Work Task Impact Search Behavior and Engagement?},
year = {2016},
isbn = {9781450337519},
publisher = {Association for Computing Machinery},
address = {New York, NY, USA},
url = {https://doi.org/10.1145/2854946.2855000},
doi = {10.1145/2854946.2855000},
booktitle = {Proceedings of the 2016 ACM on Conference on Human Information Interaction and Retrieval},
pages = {249–252},
numpages = {4},
location = {Carrboro, North Carolina, USA},
series = {CHIIR '16}
}

@inproceedings{Liu2021,
author = {Liu, Jiqun and Jung, Yong Ju},
title = {Interest Development, Knowledge Learning, and Interactive IR: Toward a State-based Approach to Search as Learning},
year = {2021},
isbn = {9781450380553},
publisher = {Association for Computing Machinery},
address = {New York, NY, USA},
url = {https://doi.org/10.1145/3406522.3446015},
doi = {10.1145/3406522.3446015},
booktitle = {Proceedings of the 2021 Conference on Human Information Interaction and Retrieval},
pages = {239–248},
numpages = {10},
location = {Canberra ACT, Australia},
series = {CHIIR '21}
}

@inbook{Huang2025,
author = {Huang, Kun and Guo, Qionghao and Zhou, Chenyu and Hao, Xijia},
title = {How Do Emotional Tasks Influence Information Seeking Behavior?},
year = {2025},
isbn = {9798400710933},
publisher = {Association for Computing Machinery},
address = {New York, NY, USA},
url = {https://doi.org/10.1145/3677389.3703185},
booktitle = {Proceedings of the 24th ACM/IEEE Joint Conference on Digital Libraries},
articleno = {73},
numpages = {3}
}

@article{Nickerson1998,
author = {Raymond S. Nickerson},
title ={Confirmation Bias: A Ubiquitous Phenomenon in Many Guises},
journal = {Review of General Psychology},
volume = {2},
number = {2},
pages = {175-220},
year = {1998},
doi = {10.1037/1089-2680.2.2.175},
URL = {https://doi.org/10.1037/1089-2680.2.2.175},
eprint = {https://doi.org/10.1037/1089-2680.2.2.175}
}

@inproceedings{Azzopardi2021,
author = {Azzopardi, Leif},
title = {Cognitive Biases in Search: A Review and Reflection of Cognitive Biases in Information Retrieval},
year = {2021},
isbn = {9781450380553},
publisher = {Association for Computing Machinery},
address = {New York, NY, USA},
url = {https://doi.org/10.1145/3406522.3446023},
doi = {10.1145/3406522.3446023},
booktitle = {Proceedings of the 2021 Conference on Human Information Interaction and Retrieval},
pages = {27–37},
numpages = {11},
location = {Canberra ACT, Australia},
series = {CHIIR '21}
}

@inproceedings{Reimer2023,
author = {Reimer, Jan Heinrich and Schmidt, Sebastian and Fr\"{o}be, Maik and Gienapp, Lukas and Scells, Harrisen and Stein, Benno and Hagen, Matthias and Potthast, Martin},
title = {The Archive Query Log: Mining Millions of Search Result Pages of Hundreds of Search Engines from 25 Years of Web Archives},
year = {2023},
isbn = {9781450394086},
publisher = {Association for Computing Machinery},
address = {New York, NY, USA},
url = {https://doi.org/10.1145/3539618.3591890},
doi = {10.1145/3539618.3591890},
booktitle = {Proceedings of the 46th International ACM SIGIR Conference on Research and Development in Information Retrieval},
pages = {2848–2860},
numpages = {13},
location = {Taipei, Taiwan},
series = {SIGIR '23}
}

@inproceedings{Rekabsaz2021,
author = {Rekabsaz, Navid and Lesota, Oleg and Schedl, Markus and Brassey, Jon and Eickhoff, Carsten},
title = {TripClick: The Log Files of a Large Health Web Search Engine},
year = {2021},
isbn = {9781450380379},
publisher = {Association for Computing Machinery},
address = {New York, NY, USA},
url = {https://doi.org/10.1145/3404835.3463242},
doi = {10.1145/3404835.3463242},
booktitle = {Proceedings of the 44th International ACM SIGIR Conference on Research and Development in Information Retrieval},
pages = {2507–2513},
numpages = {7},
location = {Virtual Event, Canada},
series = {SIGIR '21}
}

@inproceedings{Craswell2020,
author = {Craswell, Nick and Campos, Daniel and Mitra, Bhaskar and Yilmaz, Emine and Billerbeck, Bodo},
title = {ORCAS: 18 Million Clicked Query-Document Pairs for Analyzing Search},
year = {2020},
isbn = {9781450368599},
publisher = {Association for Computing Machinery},
address = {New York, NY, USA},
url = {https://doi.org/10.1145/3340531.3412779},
doi = {10.1145/3340531.3412779},
booktitle = {Proceedings of the 29th ACM International Conference on Information \& Knowledge Management},
pages = {2983–2989},
numpages = {7},
location = {Virtual Event, Ireland},
series = {CIKM '20}
}

@inproceedings{Serdyukov2014,
author = {Serdyukov, Pavel and Dupret, Georges and Craswell, Nick},
title = {Log-based personalization: the 4th web search click data (WSCD) workshop},
year = {2014},
isbn = {9781450323512},
publisher = {Association for Computing Machinery},
address = {New York, NY, USA},
url = {https://doi.org/10.1145/2556195.2556207},
doi = {10.1145/2556195.2556207},
booktitle = {Proceedings of the 7th ACM International Conference on Web Search and Data Mining},
pages = {685–686},
numpages = {2},
location = {New York, New York, USA},
series = {WSDM '14}
}

@misc{zou2022,
      title={A Large Scale Search Dataset for Unbiased Learning to Rank}, 
      author={Lixin Zou and Haitao Mao and Xiaokai Chu and Jiliang Tang and Wenwen Ye and Shuaiqiang Wang and Dawei Yin},
      year={2022},
      eprint={2207.03051},
      archivePrefix={arXiv},
      primaryClass={cs.AI},
      url={https://arxiv.org/abs/2207.03051}, 
}

@inproceedings{DBLP:conf/infoscale/PassCT06,
  author       = {Greg Pass and
                  Abdur Chowdhury and
                  Cayley Torgeson},
  editor       = {Xiaohua Jia},
  title        = {A picture of search},
  booktitle    = {Proceedings of the 1st International Conference on Scalable Information
                  Systems, Infoscale 2006, Hong Kong, May 30-June 1, 2006},
  series       = {{ACM} International Conference Proceeding Series},
  volume       = {152},
  pages        = {1},
  publisher    = {{ACM}},
  year         = {2006},
  url          = {https://doi.org/10.1145/1146847.1146848},
  doi          = {10.1145/1146847.1146848},
  bibsource    = {dblp computer science bibliography, https://dblp.org}
}

@Inbook{Song2021,
author="Song, Ruihua
and Zhang, Min
and Luo, Cheng
and Sakai, Tetsuya
and Liu, Yiqun
and Dou, Zhicheng",
editor="Sakai, Tetsuya
and Oard, Douglas W.
and Kando, Noriko",
title="SogouQ: The First Large-Scale Test Collection with Click Streams Used in a Shared-Task Evaluation",
bookTitle="Evaluating Information Retrieval and Access Tasks: NTCIR's Legacy of Research Impact",
year="2021",
publisher="Springer Singapore",
address="Singapore",
pages="143--150",
isbn="978-981-15-5554-1",
doi="10.1007/978-981-15-5554-1_10",
url="https://doi.org/10.1007/978-981-15-5554-1_10"
}

@inproceedings{DBLP:conf/cikm/ChenMLZM19,
  author       = {Jia Chen and
                  Jiaxin Mao and
                  Yiqun Liu and
                  Min Zhang and
                  Shaoping Ma},
  editor       = {Wenwu Zhu and
                  Dacheng Tao and
                  Xueqi Cheng and
                  Peng Cui and
                  Elke A. Rundensteiner and
                  David Carmel and
                  Qi He and
                  Jeffrey Xu Yu},
  title        = {TianGong-ST: {A} New Dataset with Large-scale Refined Real-world Web
                  Search Sessions},
  booktitle    = {Proceedings of the 28th {ACM} International Conference on Information
                  and Knowledge Management, {CIKM} 2019, Beijing, China, November 3-7,
                  2019},
  pages        = {2485--2488},
  publisher    = {{ACM}},
  year         = {2019},
  url          = {https://doi.org/10.1145/3357384.3358158},
  doi          = {10.1145/3357384.3358158},
  bibsource    = {dblp computer science bibliography, https://dblp.org}
}

@inproceedings{DBLP:conf/ercimdl/MayrK17,
  author       = {Philipp Mayr and
                  Ameni Kacem},
  editor       = {Jaap Kamps and
                  Giannis Tsakonas and
                  Yannis Manolopoulos and
                  Lazaros S. Iliadis and
                  Ioannis Karydis},
  title        = {A Complete Year of User Retrieval Sessions in a Social Sciences Academic
                  Search Engine},
  booktitle    = {Research and Advanced Technology for Digital Libraries - 21st International
                  Conference on Theory and Practice of Digital Libraries, {TPDL} 2017,
                  Thessaloniki, Greece, September 18-21, 2017, Proceedings},
  series       = {Lecture Notes in Computer Science},
  volume       = {10450},
  pages        = {560--565},
  publisher    = {Springer},
  year         = {2017},
  url          = {https://doi.org/10.1007/978-3-319-67008-9\_46},
  doi          = {10.1007/978-3-319-67008-9\_46},
  bibsource    = {dblp computer science bibliography, https://dblp.org}
}

\end{document}